\documentclass[final,3p,times,twocolumn]{elsarticle}
 \biboptions{comma,sort&compress}
\usepackage{here}
 \usepackage{graphicx}
  \usepackage{epsfig}

\def\nin{\noindent}
\def\beq{\begin{equation}}
\def\eeq{\end{equation}}
\def\bea{\begin{eqnarray}}
\def\eea{\end{eqnarray}}

\journal{Nuc. Phys. (Proc. Suppl.)}

\begin{document}

\begin{frontmatter}



\title{Search for the ``Dark Photon'' and  the ``Dark Higgs'' at Belle} 

 \author{Igal Jaegle for the Belle Collaboration}
  \address{Department of Physics and Astronomy, University of Hawai'i
 at Manoa,\\ 2505 Correa Road, Honolulu 96822, Hawai'i, USA.}
\ead{igjaegle@gmail.com}


\begin{abstract}
\noindent
The expected sensitivity of Belle is reported for the Dark Photon, $A$, and 
Dark Higgs, $h'$ searches, for mass ranges, respectively of 0.27 $< m_A <$ 
3 GeV/c$^2$ and 0.54 $< m_{h'} <$ 10 GeV/c$^2$. The Dark Photon and 
Dark Higgs will be searched for in the Higgs-strahlung channels: 
$e^+e^- \rightarrow Ah'$, with $h' \rightarrow AA$ and 
$A \rightarrow l^+l^-$ (with $l$ = $e$ or $\mu$). Results will be unblinded soon.

\end{abstract}


\end{frontmatter}


\section{Introduction}
\nin
Ordinary matter represents 4 $\%$ of the total energy budget of the universe.The remainder is believed to be partitioned as either Dark 
Energy (73$\%$) or Dark Matter (23$\%$), but the nature of those components is unknown. However, experimental results from direct Dark Matter 
 searches, (e.g. DAMA/LIBRA) \cite{DAMA,CDMS-II,XENON,CoGeNT} and other experimental observations that may be interpreted as
deviations from the Standard Model (e.g. g-2 \cite{g2}), can be explained in the Dark Matter (DM) models by the inclusion of an additional interaction, a dark U(1) interaction \cite{Pospelov2007,Arkani2008,Chun2008,Cheung2009,Katz2009,Morrissey2009,Goodsell2009,Baumgart2009,Nomura2008,Alves2009,Jaeckel2010}.  This interaction, which is mediated by a dark U(1) boson, also known as
 the ``Dark Photon'', typically has very small couplings to Standard Model particles.

Dark gauge bosons are postulated to have low mass; of order MeV to GeV due to astrophysical constraints 
\cite{PAMELA,Fermi}. These astrophysical observations include: excesses in the cosmic-ray flux of electrons and/or positrons above expected background beyond normal astrophysical processes and the expected flux of protons and/or anti-protons. Dark matter could be charged under the dark U(1) symmetry group and then the observed excess might correspond to dark matter annihilating into a Dark Photon $A$, which in turn decays into $l^+l^-$ (with $l$ = $e$ or $\mu$ or possibly $\tau$ if energetically allowed).

 The ideal tools to discover such particles are therefore not the highest energy hadron collider experiments, 
but lower-energy lepton high-luminosity collider experiments such as Belle/BelleII and BaBar/SuperB, or dedicated fixed target
 experiments, several of which are planned or already under construction at JLAB (Newport News, USA) or at MAMI 
(Mainz, Germany), for example. In Belle, work on dark gauge boson searches was started only recently, and has focused on
 the strategies proposed by \cite{Reece2009,Essig2009,Batell2009,Bossi2009,Yin2009}. These proceedings will focus on the so-called
Higgs-strahlung channel, $e^+e^- \rightarrow Ah'$ and in particular the decay mode with $3e^+3e^-$ and $3\mu^+3\mu^-$ final states. The dark U(1) symmetry group could be spontaneously broken, often by 
a Higgs mechanism, adding a dark Higgs $h'$ (or dark Higgses) to these models.  The dark photon $A$ can decay into either $l^+l^-$, hadrons or 
invisible particles and $h'$ into $AA$, $l^+l^-$, hadrons or invisible particles. The decay mode of the $A$ and $h'$ depends of the mass difference between $A$ and $h'$ \cite{Batell2009}: (a) $m_{h'} < m_A$: $h' \rightarrow$ invisible particles, (b) $m_A < m_{h'} < 2m_A$: $h' \rightarrow l^+l^-$ or hadrons, (c) $m_{h'} > 2m_A$: $h' \rightarrow AA$. Case (c) will be discussed 
in this proceedings.

The $A$ and $h'$ do not necessarily have prompt decays \cite{Batell2009,Essig2010}. The decay length of the dark photon is a function of the dark photon coupling strength to Standard Model fermions
 and is proportional to the inverse of the square of the dark photon coupling. The decay length of the dark Higgs varies for $m_{h'} > 2m_A$ between being prompt and one meter. In the Higgs-strahlung channel, two couplings are involved: the  electromagnetic coupling of the dark photon to the Standard Model fermions, $\alpha'$; and the dark photon coupling to the dark Higgs, $\alpha_D$. KLOE and BaBar reported their searches on the dark photon and the dark Higgs \cite{KLOE} and \cite{BaBar} (see also contribution of B. Oberhof to these proceedings \cite{Oberhof}). KLOE focused their search on $m_{h'} < m_A$ and BaBar on $m_{h'} > 2m_A$ (and with $A$ and $h'$ prompt), but no signal was found. Only BaBar could set limits for the highest mass range,  $0.8 < m_{h'} < 10.0$ GeV/c$^2$ and $0.25 < m_{A} < 3.0$ GeV/c$^2$. BaBar looked at two types of final states, fully reconstructed ($3l^+l^-$, $2l^+2l^-\pi^+\pi^-$ and $2\pi^+2\pi^-l^+l^-$ where $l=e,\mu$); and partially reconstructed ($2\mu^+2\mu^-X$  and $e^+e^-\mu^+\mu^-X$, where $X$ denotes any final state other than a pair of pions or leptons). The reason to look for partially reconstructed final states is that above $m_A>$ 1 GeV/c$^2$, final states with hadrons dominate assuming $BF(A\rightarrow e^+e^-) + BF(A\rightarrow \mu^+\mu^-) + BF(A\rightarrow hadrons) = 1$ and that  $\frac{BF(A\rightarrow hadrons)}{BF(A\rightarrow \mu^+\mu^-)} = R$, where $R$ is the hadronic cross section ratio, $R =\frac{\sigma(e^+e^- \rightarrow hadrons)}{\sigma(e^+e^- \rightarrow \mu^+\mu^-)}$. BaBar reported six candidate events detected: one $4\mu2\pi$, two $4\pi2\mu$, two $4\pi2e$ and one $4\mu X$ events, in a $\sim$ 500 fb$-1$ data set. There were no candidates in the final states with six leptons. The number of events detected were consistent with background expectations. For low mass range of the dark photon: $1 < m_{A} < 300$ MeV/c$^2$, part of the space, coupling strength to Standard Model fermions versus dark photon mass, have been excluded by Beam dump experiments (\cite{Blumlein:1991xh,Blumlein:1990ay,Barabash:1992uq,Blumlein2011,E137,E141,E774}) by looking at reaction $pp \rightarrow A X$ for example.


\section{Experimental setup}
The Belle detector is a large-solid-angle magnetic spectrometer, which consists of a silicon vertex detector (SVD), a central drift chamber (CDD), 
an array of aerogel threshold Cerenkov counters (ACC), a barrel-like arrangement of time-of-flight scintillation counters (TOF), and an electromagnetic
 calorimeter (ECL) of CsI(Tl) crystals located inside a super-conducting solenoid that provides a 1.5 T 
magnetic field. An iron flux-return (KLM) located outside the coil is optimized to detect K$^0_L$ mesons and to identify muons. A detailed description can be found in \cite{setup}.  
Belle is currently being upgraded to Belle II, an upgraded detector for operation at SuperKEKB, which will have 40 times higher luminosity than KEKB \cite{tdr}. The KEKB collider \cite{kek}, located in Tsukuba, Japan, is the world's highest-luminosity electron-position collider. KEKB has produced more than one $\mathrm{ab}^{-1}$ of data at center-of-mass energies
 corresponding to the $\Upsilon$(1S) to $\Upsilon$(5S) resonances, and in the nearby continuum.
\nin


\section{Particle and reaction identification}
The sub-systems used to identify the electrons are primarily the ECL, which measured the energy and the CDC, which measured the momentum \cite{Hanagaki}.
 The muons are identified by using the KLM and an analysis combining the measurements of penetration depth, the charged track, and the muon cluster matching \cite{Abashian}. Events with six charged tracks with three pairs of opposite charges are considered for analysis. Furthermore, in order to maintain a high detection efficiency we require that at least three out of the six tracks be identified as leptons.
\nin

\section{Analysis strategy}
\nin

Events with six lepton final states from  $e^+e^-\rightarrow Ah'\rightarrow AAA \rightarrow 3l^+3l^-$ ($l$ = $e$ or $\mu$) are reconstructed. Energy and momentum conservation is
required. The invariant mass for each combination of leptons is required to be consistent with three distinct $A \rightarrow l^+l^-$. Combinations with three ``equal'' masses ($m^1_{ll}$, $m^2_{ll}$ and $m^3_{ll}$)  and $m_{llll} > 2m_{ll}$ are kept. The ``equality'' is defined as follows: $ m^{mean}_{ll} - 3.\sigma(m^{mean}_{ll}) < m^{1,2,3}_{ll} < m_{mean} + 3.\sigma(m^{mean}_{ll})$, with $m^{mean}_{ll}$ the mean mass of the three dark photon candidates and $\sigma$ is the width of the signal as function of the dark mass which is taken from Monte Carlo (MC) simulation. The detection efficiency of Belle was modeled with MC simulations based on the GEANT4 package \cite{GEANT4}. The simulation includes all relevant properties of the sub-systems, including geometrical
acceptance, charged particle identification, trigger efficiency, response of all detector modules, and selection criteria. The MC also includes information about inefficient individual detector modules. 
Belle can achieve, on average, a detection efficiency of 20 $\%$ and 40 $\%$ respectively for 6 electron and 6 muon final states.
%

\section{Background estimation}
\begin{figure}[hbt] 
\centerline{\includegraphics[width=6.cm]{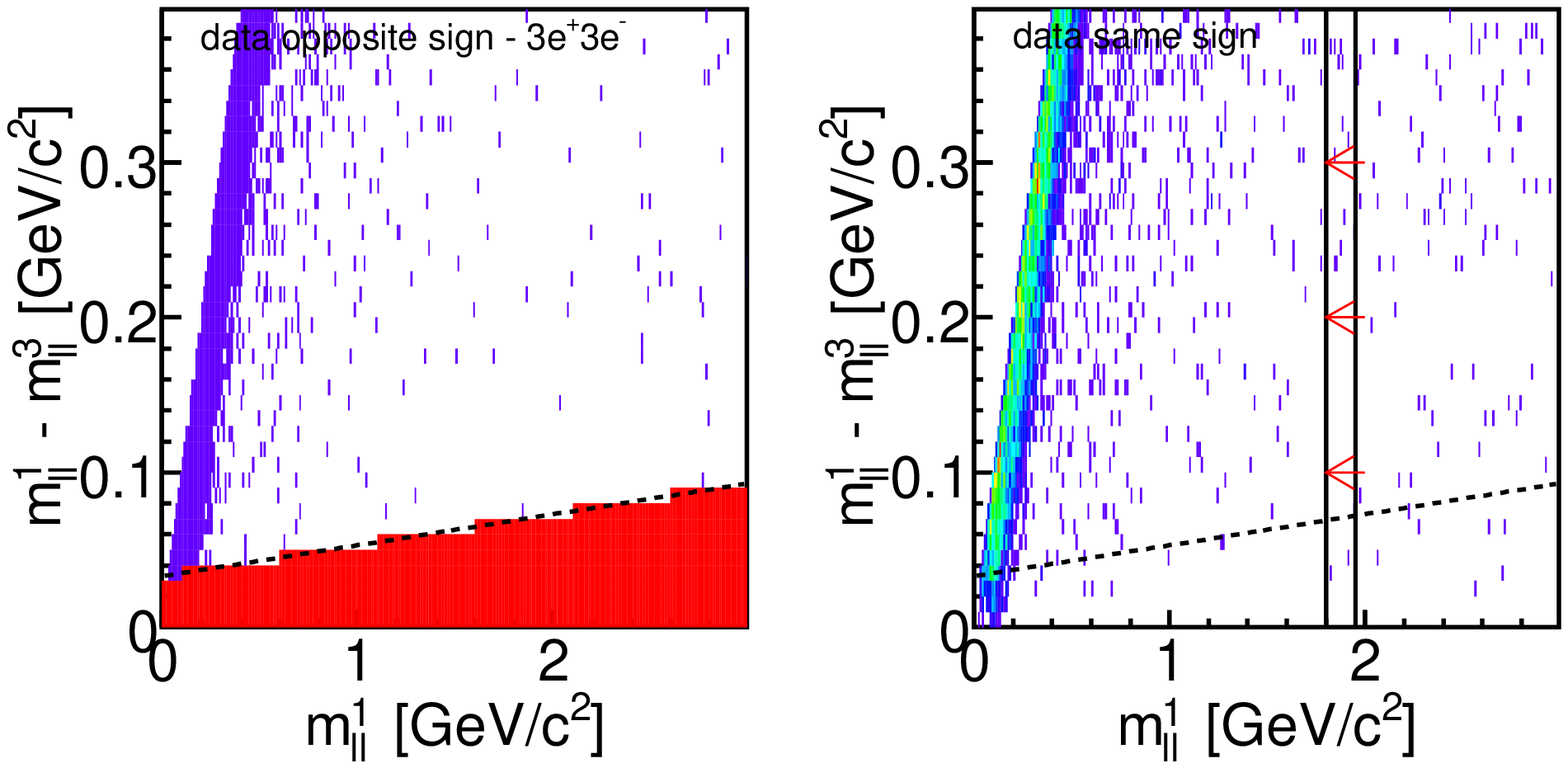}}
\centerline{\includegraphics[width=6.cm]{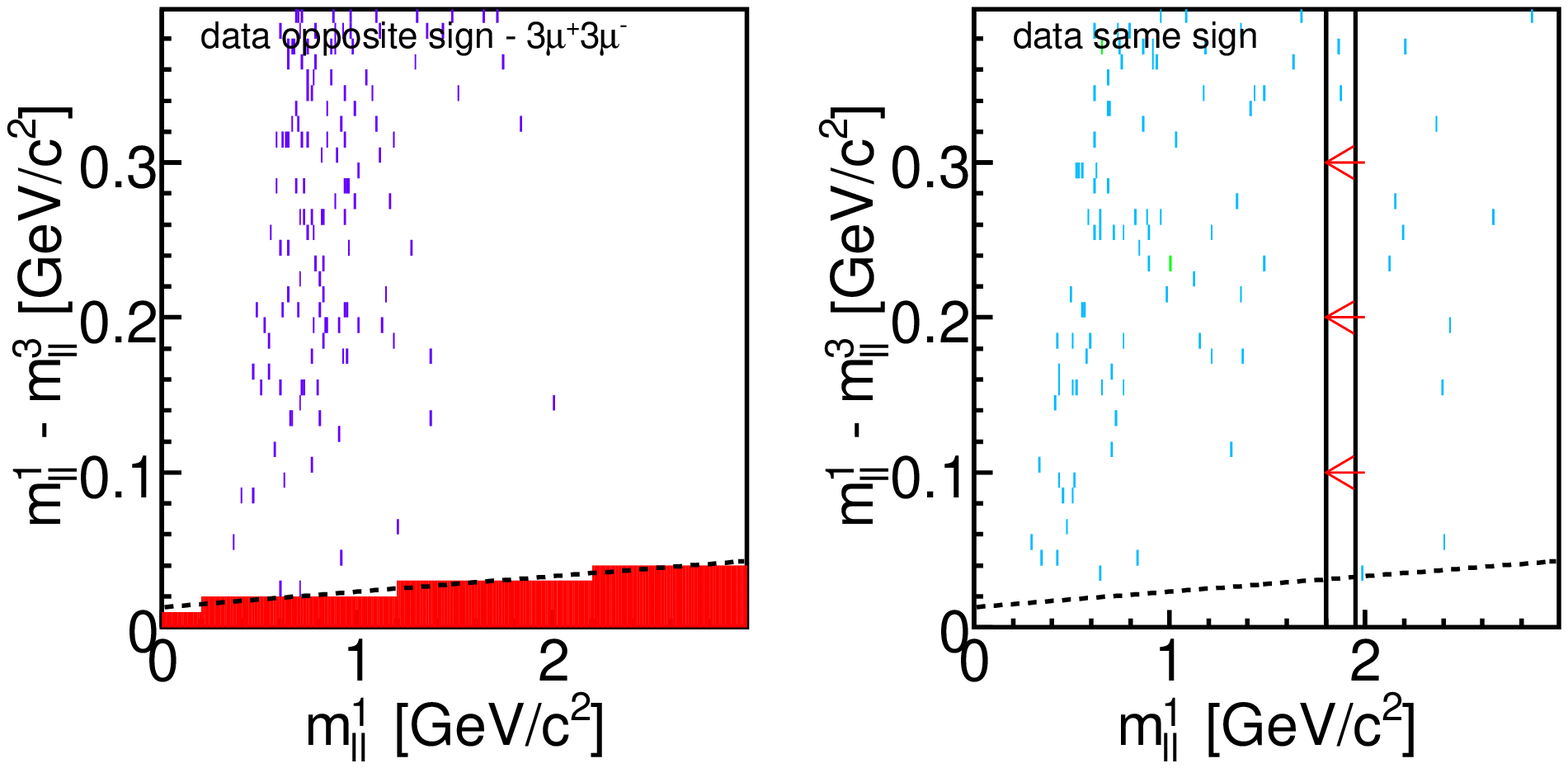}}
\caption{\scriptsize $m_{ll}^1$ versus $m_{ll}^1-m_{ll}^3$ for 6e final state (top) and 6$\mu$ final state (bottom).  Top-left and
bottom-left, opposite sign: $(e^+e^-)(e^+e^-)(e^+e^-)$. Top-right and bottom-right, same sign:  $(e^+e^-)(e^+e^+)(e^-e^-)$. The
 2D histograms are divided into 20 slices, each slice is then projected on the $m_{ll}^1-m_{ll}^3-$axis. The red arrow (right) shows 
slice 10, corresponding to $m_{ll}^1$ = 1.9 $\pm$ 0.1 GeV/c$^2$ and the direction of the projection. The red box on the left (top and bottom) corresponds to the blinded signal region.}
\label{fig3} 
\end{figure}

\nin
The background estimation is based on a data driven method. In this method, all combinations that have two pairs where the leptons are combined with their wrong-sign partner and one pair with opposite charge, $(l^-l^-)(l^+l^+)(l^+l^-)$, are kept. The three masses are ordered in decreasing order: $m_{ll}^1 > m_{ll}^2 > m_{ll}^3$.  The mass difference between the $m_{ll}^1$, the highest
 mass, and $m_{ll}^3$ the lowest mass: $m_{ll}^1 - m_{ll}^3$ is then calculated. Figure~\ref{fig3} shows the mass difference $m_{ll}^1 - m_{ll}^3$ as function of the mass $m_{ll}^1$ for the 6 
electrons (Figure~\ref{fig3}-top) and 6 muons (Figure~\ref{fig3}-bottom) final states. For conciseness, the following notations are used: ``same sign'' for $(l^+l^-)(l^+l^+)(l^-l^-)$ and ``opposite sign'' for
  $(l^+l^-)(l^+l^-)(l^+l^-)$ (with $l$ = $e$ or $\mu$). The scatter plot for the opposite sign: $(e^+e^-)(e^+e^-)(e^+e^-)$ - Figure~\ref{fig3}-top-right and 
$(\mu^+\mu^-)(\mu^+\mu^-)(\mu^+\mu^-)$ - Figure~\ref{fig3}-bottom-right have their signal region blinded (red bands), since a blind analysis technique is used and not all selection criteria have been validated.
The signal region for the same sign scatter plots
($(e^+e^-)(e^+e^+)(e^-e^-)$ - Figure~\ref{fig3}-top-left and $(\mu^+\mu^-)(\mu^+\mu^+)(\mu^-\mu^-)$ - Figure~\ref{fig3}-bottom-left) is unblinded. 

\begin{figure}[hbt] 
\centerline{\includegraphics[width=5cm]{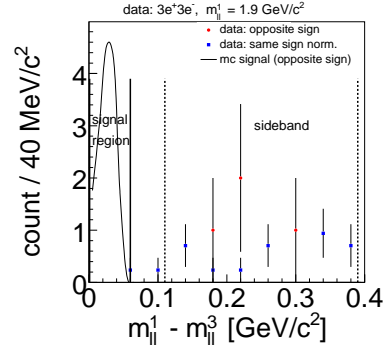}}
\caption{\scriptsize Projection on $m_{ll}^1-m_{ll}^3$ for $m_{ll}^1$ = 1.9 $\pm$ 0.1 GeV/c$^2$. The ``same sign''-$(e^+e^-)(e^+e^+)(e^-e^-)$  (black point) distribution has been normalized to the ``opposite sign''- $(e^+e^-)(e^+e^-)(e^+e^-)$ (red point) distribution using the side band area.}
\label{fig4} 
\end{figure}

\nin
The background is then estimated for different $m^1_{ll}$ mass regions as illustrated for the $m^1_{ll}$ = 1.9 $\pm$ 0.1 GeV/$c^2$ band by Figure~\ref{fig4}. It is assumed that the backgrounds have the same shape in the side band region of  the same sign distribution and the opposite sign distribution but not necessarily the same number of events. Therefore, the same sign distribution is normalized to the opposite sign distribution by a factor calculated for each  $m^1_{ll}$. The expected background is then the scaled number of events counted in the signal region of the same sign distribution. The signal obtained in the MC simulation is shown as a black curve in Figure~\ref{fig3} (for the opposite sign).
\nin

\nin
\nin

\section{Expected sensitivity}

\begin{figure}
\includegraphics[width=3.5cm]{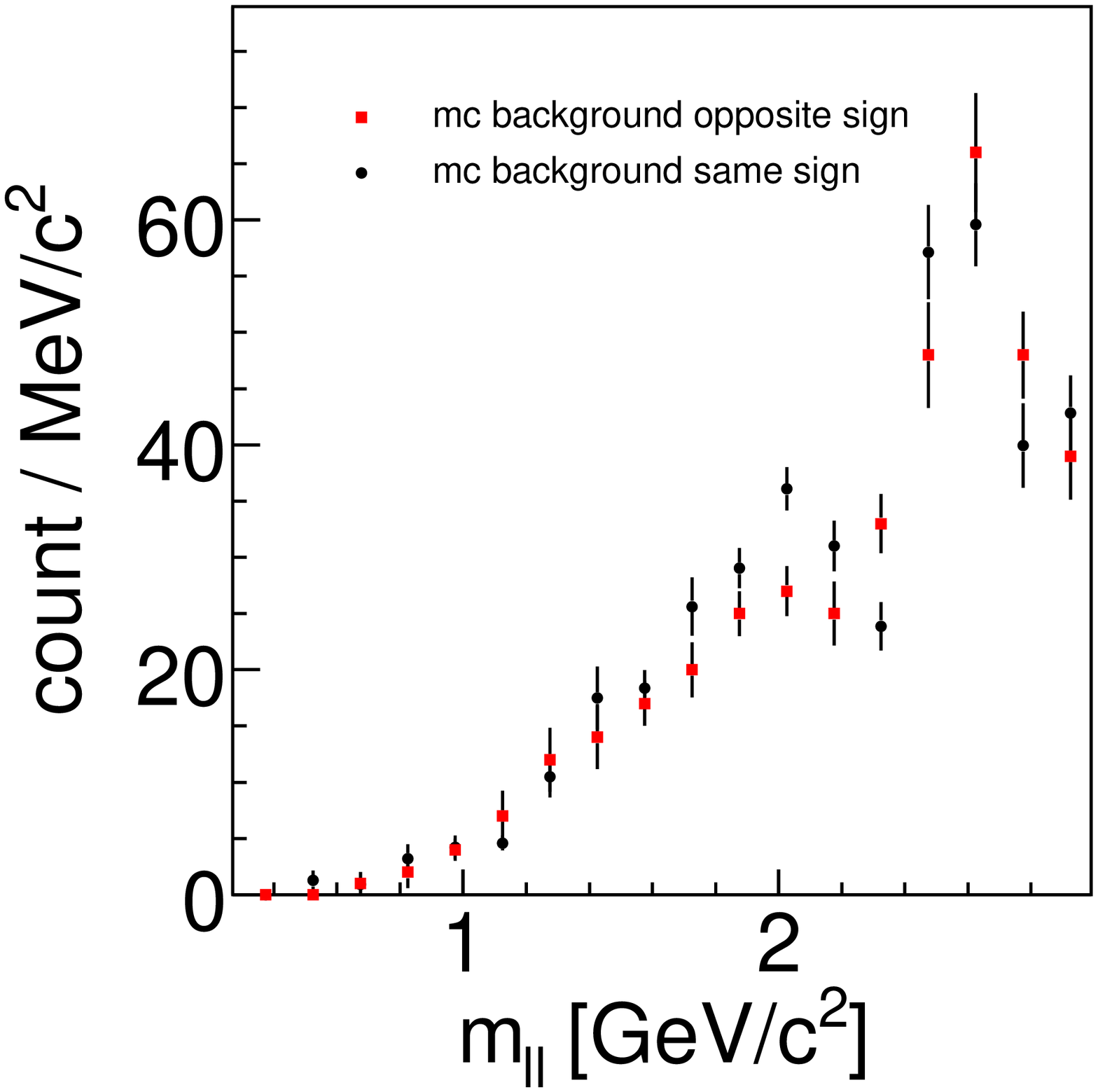}
\includegraphics[width=3.5cm]{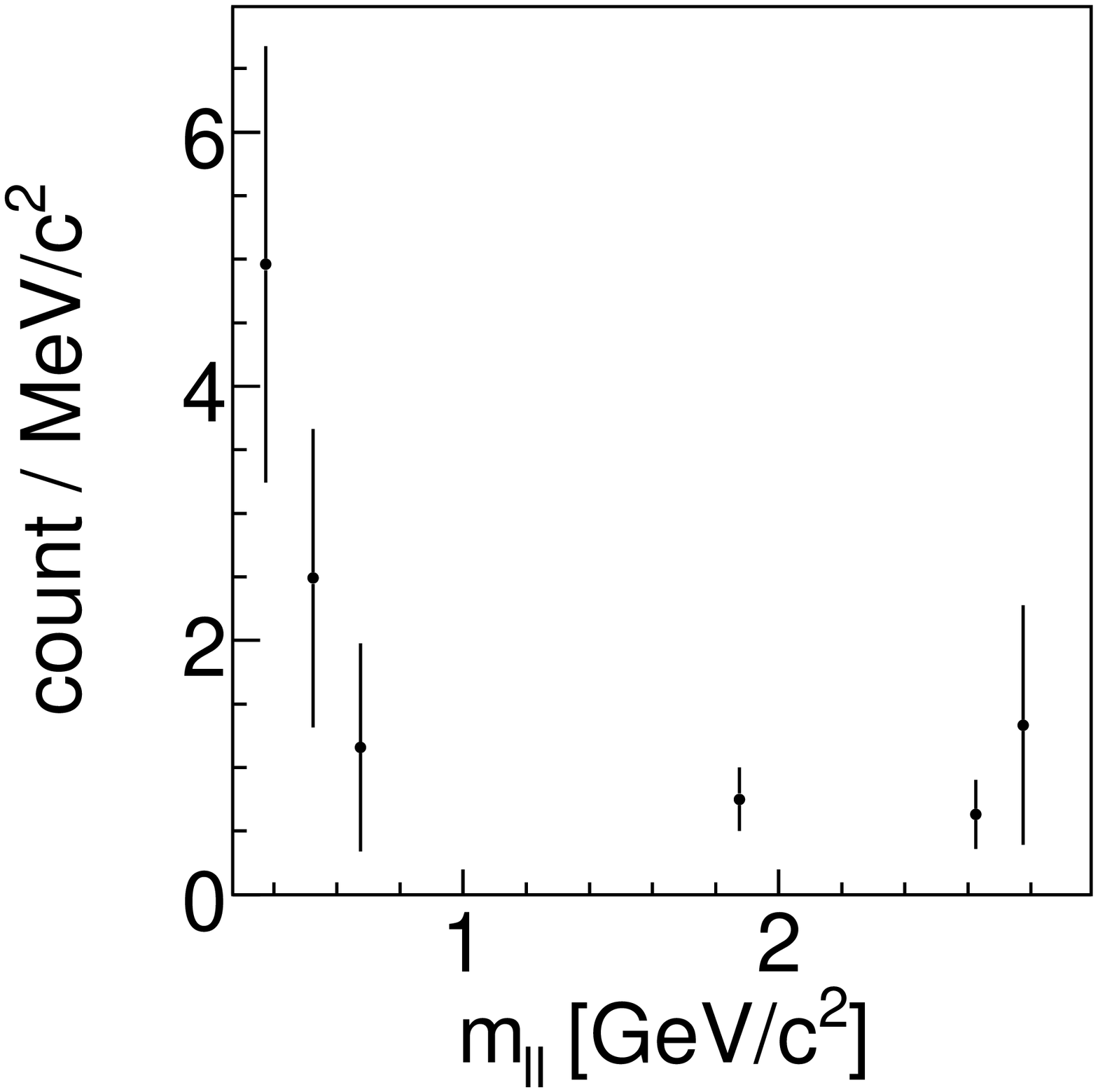}
\caption{\scriptsize Left, background estimation method verified successfully with MC 
Right, for experimental data, predicted BG is $>$ 20 events for the electron final state.}
\label{fig5} 
\end{figure} 
\nin

The expected background with no constraint on the impact parameters and the vertex positions is less than 25 events and less than 5 events with a constraint on the impact parameters but no constraint on the vertex positions for the six electron final states. No events are found in either case for the six muon final state. The case with no constraint on the impact parameter and on the vertex position corresponds to 
displaced vertex positions up to 80 cm from the interaction point and according to \cite{Batell2009,Essig2010} up to $\epsilon \sim  10^{-6}$.  Here, $\epsilon = {\frac{\alpha'}{\alpha}}^2$ is the strength of the dark photon mixing with the Standard Model hypercharge gauge
boson and $\alpha$, the electromagnetic coupling. For the case with a constraint on the impact parameters, the decay length can go up to 6 mm i.e. up to $\epsilon \sim 10^{-4}$.

Figure~\ref{fig5}-left shows a MC simulation test of the background estimation method. The interaction $e^+e^- \rightarrow 3e^+3e^-$ have been generated in phase space. The background estimated from the same sign distribution is consistent with the number of events counted in the signal region of the opposite sign for all $m_A$ candidates. Figure~\ref{fig5}-right shows the expected background deduced from the data same sign analysis.  

\begin{figure}
\includegraphics[width=3.5cm]{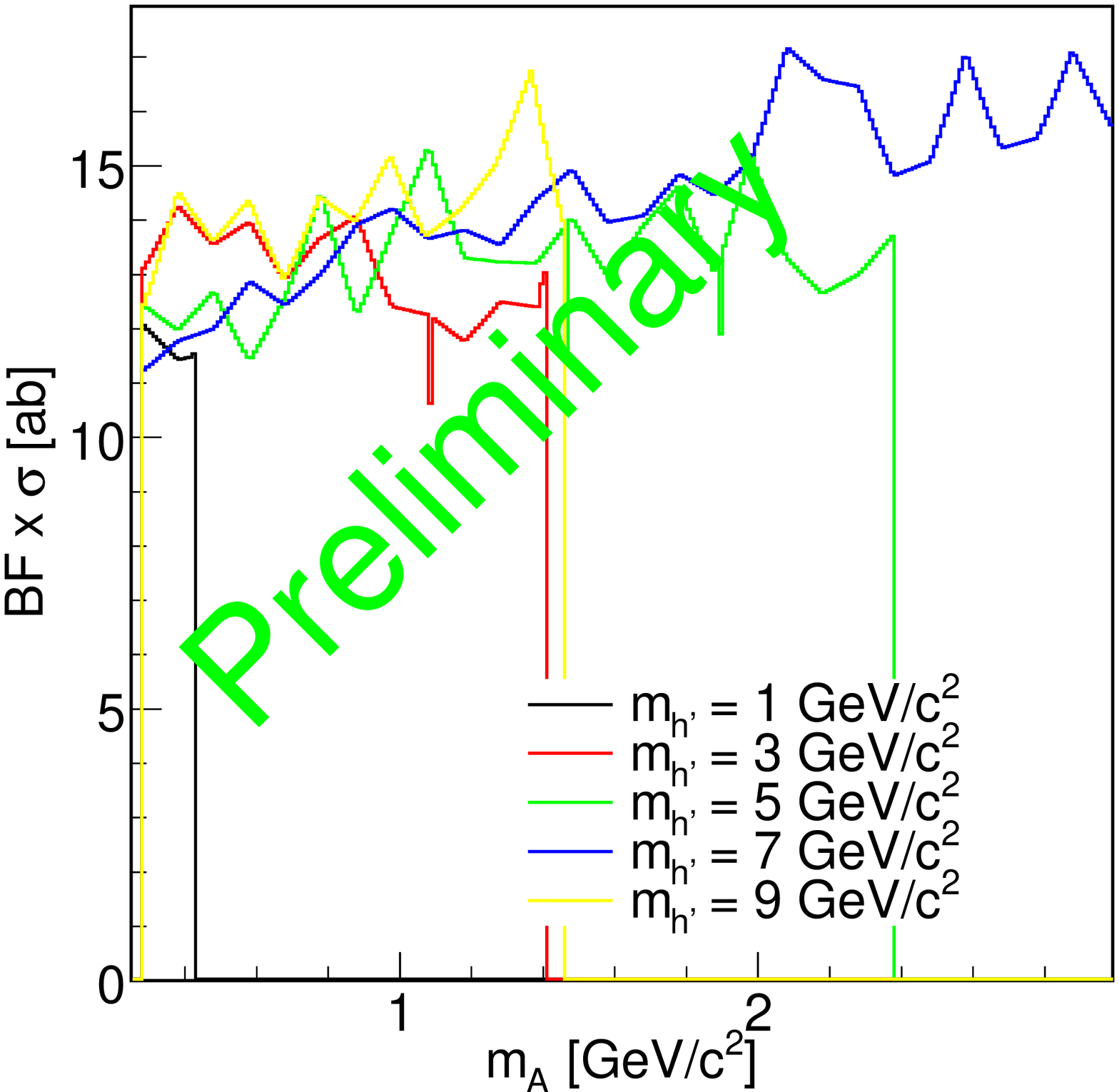}
\includegraphics[width=3.5cm]{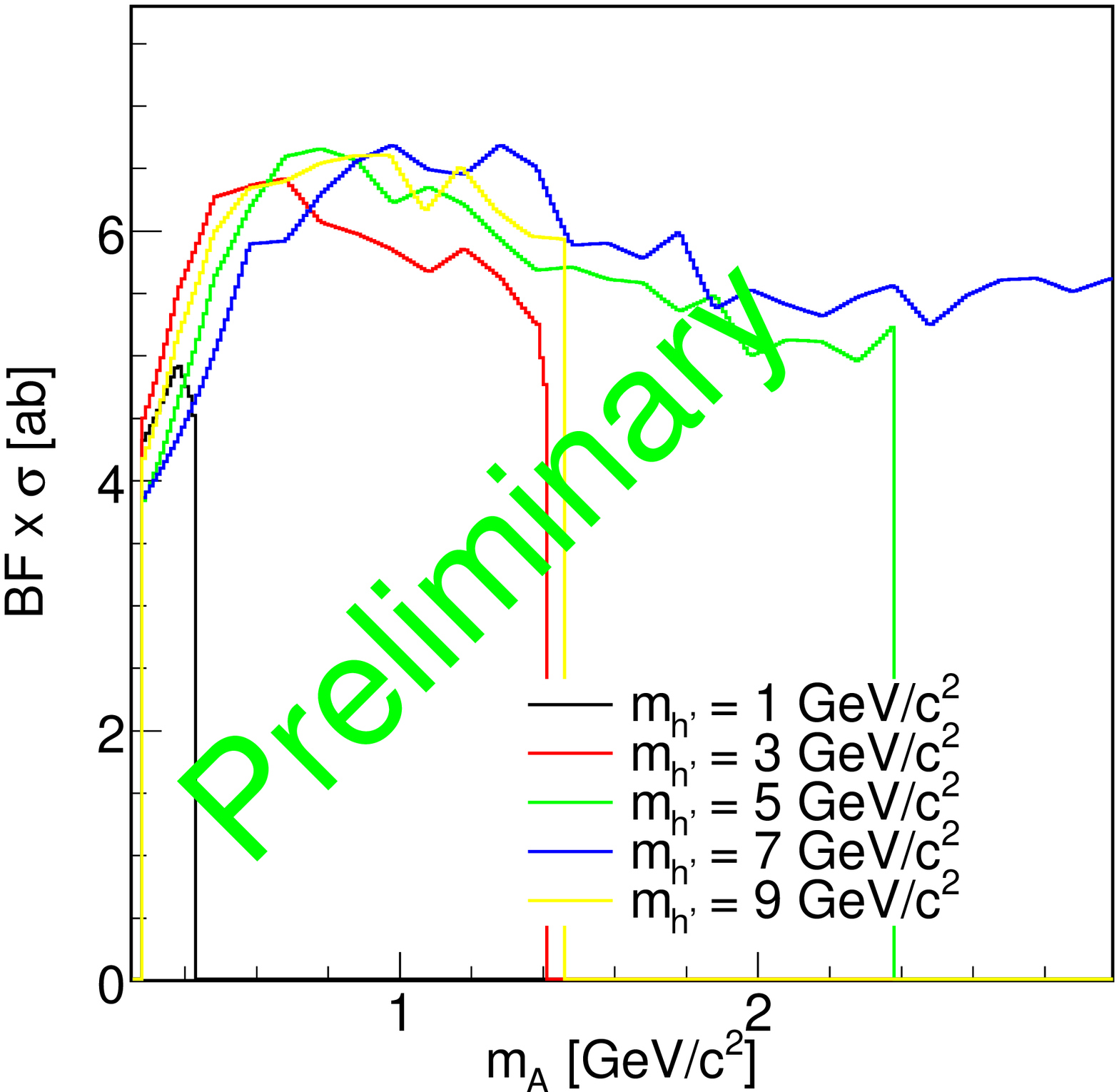}
\caption{\scriptsize Sensitivity as function of the ``Dark Photon'' mass for different ``Dark Higgs'' mass hypothesis. Left: 6e. Right: $6\mu$.}
\label{fig6} 
\end{figure}
\begin{figure}
\includegraphics[width=3.5cm]{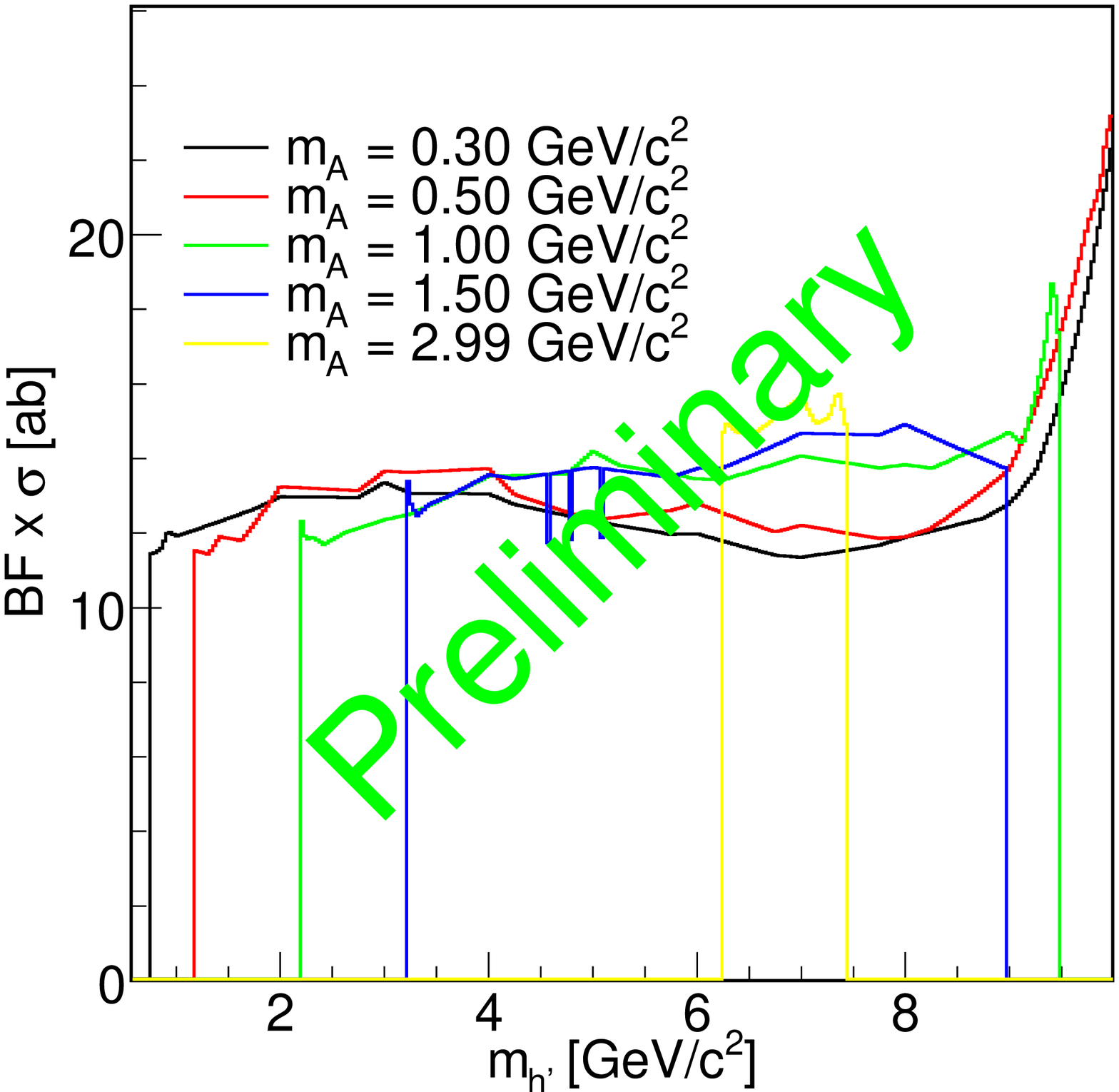}
\includegraphics[width=3.5cm]{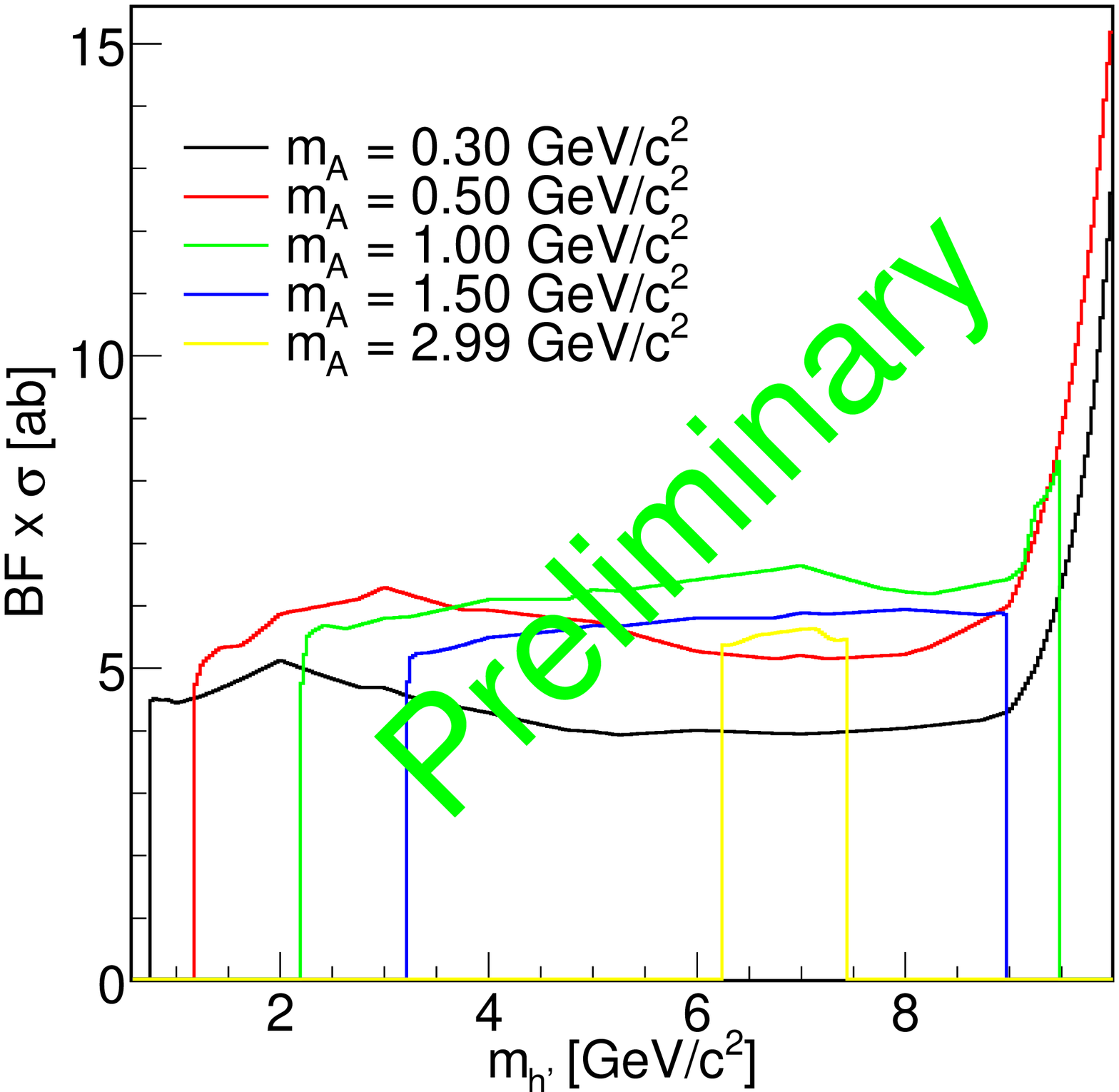}
\caption{\scriptsize Sensitivity as function of the ``Dark Higgs'' mass for different ``Dark Photon'' mass hypothesis. Left: 6e. Right: $6\mu$.}
\label{fig7} 
\end{figure}

A statistical method based on the Feldman-Cousins approach \cite{FeldCou} is used to then calculate preliminary upper limits (90$\%$ CL) for a number of ``observed'' events equal to the estimated background events for the full luminosity and the prompt case. Figure~\ref{fig6} (left - six electron, right six muons final states) shows the preliminary sensitivity limit for different ``Dark Higgs'' mass hypotheses. Figure~\ref{fig7} (left - six electron, right 6 muons final states) shows the preliminary sensitivity limit for different ``Dark Photon'' mass hypothesis. Due to the expected low level of background the sensitivity scales nearly linearly with the integrated luminosity. 

\nin
\nin

\section{Conclusions}
\nin
The Dark Photon and the Dark Higgs are searched for in the mass ranges: $0.27 < m_A < 3$ GeV/c$^2$ and  $0.54 < m_{h'} < 10.86$ GeV/c$^2$. Based on control data samples, it was found that the background is small, implying that the detection sensitivity scales nearly linearly with integrated luminosity. The expected preliminary Belle sensitivities have been shown. Results for data will be unblinded soon.

\section*{Acknowledgements}
\nin
The authors would like to thank the organizers (LPTA-Montpellier) of  the 16th International Conference in Quantum ChromoDynamics 2-6th July 2012 (Montpellier-France) for their kind invitation and hospitality and congratulate them for a successful workshop. We acknowledge support from the U.S.Department of Energy under Award
Number DE-SC0007852.






\begin{thebibliography}{999}
\vspace*{-0.25cm}

\bibitem{DAMA} DAMA Collaboratopn, R. Bernabei, et al. Eur. Phys. J. C 56 (2008).
\bibitem{CDMS-II} CDMS-II Collaboration, Z. Ahmed et al., Published Feb 11, 2010, Science 327:1619-1621 arXiv:0912.3592 (2010).
\bibitem{XENON} XENON Collaboration, E. Aprile, et al. Phys. Rev. Lett. 105:131302 (2010).
\bibitem{CoGeNT} CoGeNT collaboration, C.E. Aalseth et al. arXiv:1002.4703 (2010). 

\bibitem{g2} Muon g-2 Collaboration, G. W. Bennett et al., Phys. Rev. D73 072003 (2006).

\bibitem{Pospelov2007} M. Pospelov, A. Ritz and M. Voloshin, arXiv:0711.4866 (2007).
\bibitem{Arkani2008} N. Arkami-Hamed, D. Finkbeiner, T. Slatyer and N. Weiner, arXiv:0810.0713 (2008).
\bibitem{Chun2008} E. J. Chun and J. C. Park, arXiv:0812.0308 (2008).
\bibitem{Cheung2009} C. Cheung, J. Ruderman, L.T. Wang, and I. Yavin, arXiv:0902.3246 (2009).
\bibitem{Katz2009} A. Katz and R. Sundrum, arXiv:0902.3271 (2009).
\bibitem{Morrissey2009} D. Morrissey, D. Poland and K. Zurek, arXiv:0904.2567 (2009).
\bibitem{Goodsell2009} M. Goodsell, J. Jaeckel, J. Redondo, and A. Ringwald, arXiv:0909.0515 (2009).
\bibitem{Baumgart2009} M. Baumgart, C. Cheung, L.-T. Wang, J. Ruderman, I. Yavin, arXiv:0901.0283 (2009).
\bibitem{Nomura2008} Y. Nomura and J. Thaler, arXiv:0810.5397 (2008).
\bibitem{Alves2009} D. Alves, S. Behbabani, P. Schuster, and J. Wacker, arXiv:0903.3945 (2009).
\bibitem{Jaeckel2010} J. Jaeckel, A. Ringwald, Ann. Rev. Nucl. Part. Sci. 60 405 (2010).

\bibitem{PAMELA}PAMELA - O. Adriani et al., Nature 458, 607-609 (2009)
\bibitem{Fermi}Fermi LAT. Collaboration, M. Ackermann et al., Phys Rev. D 82, 092004 (2010)

\bibitem{Reece2009} M. Reece and L. T. Wang, JHEP 0907, 051 (2009)
\bibitem{Essig2009} R. Essig, P. Schuster, and N. Toro, arXiv:0903.3941 (2009).
\bibitem{Batell2009} B. Batell, M. Pospelov, and A. Ritz, arXiv:0903.0363 (2009).
\bibitem{Bossi2009} F. Bossi, arXiv:0904.3815 (2009).
\bibitem{Yin2009} P.-f. Yin, J. Liu, and S.-h. Zhu, arXiv:0904.4644 (2009).

\bibitem{Essig2010} R. Essig, et al. arXiv:1008.0636v1 (2010).

\bibitem{KLOE} Simona Giovannella, J. Phys.: Conf. Ser. 335 012067 (2011).
\bibitem{BaBar} J. P. Lees et al (BaBar Collaboration), Phys. Rev. Lett. 108, 211801 (2012).
\bibitem{Oberhof}B. Oberhof (BaBar Collaboration), arXiv:1209.2666 (2012)

\bibitem{Blumlein:1991xh}
  J.~Blumlein, {\it et al.},
  Int.\ J.\ Mod.\ Phys.\ A {\bf 7}  3835 (1992).


\bibitem{Blumlein:1990ay}
  J.~Blumlein, {\it et al.},
  Z.\ Phys.\ C {\bf 51} 341 (1991).


\bibitem{Barabash:1992uq}
  L.~Barabash, {\it et al.},
  Phys.\ Lett.\ B {\bf 295}  154 (1992).

\bibitem{Blumlein2011}
 J.~Blumlein and J.~Brunner, arXiv:1104.2747 (2011).

\bibitem{E137}
J. D. Bjorken, S. Ecklund, W. R. Nelson, A. Abashian, C. Church, B. Lu, L. W. Mo,
T. A. Nunamaker et al., E137 collaboration, Phys. Rev. D38 (1988) 3375.

\bibitem{E141}
E. M. Riordan, M. W. Krasny, K. Lang, P. De Barbaro, A. Bodek, S. Dasu, N. Varelas,
X. Wang et al., E141 collaboration, Phys. Rev. Lett. 59 (1987) 755.

\bibitem{E774}
A. Bross, M. Crisler, S. H. Pordes, J. Volk, S. Errede, J. Wrbanek, E774 collaboration,
Phys. Rev. Lett. 67 (1991) 2942.

\bibitem{setup} A. Abashian et al. (Belle Collaboration), Nucl. Instrum. Methods Phys. Res., Sect. A 479, 117 (2002).
\bibitem{tdr}K. Abe et al. (Belle II Collaboration), “Belle II Technical Design Report”, KEK Report 2010-1, arXiv:1011.0352v1 (2010)
\bibitem{kek}S. Kurokawa, Nucl. Intr. and Meth. A499, 1 (2003).

\bibitem{Hanagaki}K. Hanagaki et al, NIM A 485, 490 (2002).
\bibitem{Abashian}A. Abashian et al, NIM A 491, 69 (2002).

\bibitem{GEANT4}R. Brun et al., GEANT, Cern/DD/ee/84-1, (1986).

\bibitem{FeldCou}G.J. Feldman and R.D. Cousins Phys. Rev. D 57, 3873–3889 (1998) 
\end{thebibliography}








\end{document}